\documentclass[12pt]{article}
\usepackage{amsmath,amssymb}
\usepackage[dvips]{epsfig}

\title{\bf Surmounting collectively oscillating bottlenecks}
\author{{\bf D. Hennig, $^{1}$ L. Schimansky-Geier, $^{1}$ and P.
 H\"anggi $^{2}$}
\\
\\
$^1$ Institut f\"ur Physik,
  Humboldt-Universit\"at zu Berlin,\\ Newton Str. 15, D-12489 Berlin\ 
\\
\\
  $^2$ Institut f\"ur Physik, Universit\"at Augsburg,\\
  Universit\"atsstr.~1, D-86135 Augsburg, Germany\\
\\}

\begin{document}

\maketitle 
\baselineskip 24 pt
%\vspace{2in} PACS numbers:
%87.-15.v, 63.20.Kr, 63.20.Ry \\

\newpage
\begin{abstract}
\noindent We study the collective escape dynamics of a chain of
coupled, weakly damped nonlinear oscillators from a metastable state
over a barrier when driven by a thermal heat bath in combination
with a weak, globally acting  periodic perturbation. Optimal
parameter choices are identified that lead to a drastic enhancement
of  escape rates as compared to a pure noise-assisted situation. We
elucidate the speed-up of escape in the driven Langevin dynamics by
showing that the time-periodic external field in combination with the thermal fluctuations triggers an instability mechanism of the
stationary homogeneous lattice state of the system. Perturbations of
the latter provided by incoherent thermal fluctuations grow because
of a parametric resonance, leading to the formation of spatially
{\it localized modes} (LMs). Remarkably, the LMs persist in spite of
continuously impacting thermal noise. The average escape time
assumes a distinct minimum by either tuning the coupling strength
and/or the driving frequency. This weak ac-driven assisted escape in
turn implies a giant speed of  the  activation rate of such
thermally driven coupled nonlinear oscillator chains.
\end{abstract}

Ever since the seminal work by Kramers (for a comprehensive review
see Ref.~\cite{RMP}) we witness a continual interest in the dynamics
of escape processes of single particles, of coupled degrees of
freedom or of chains of coupled objects out of metastable states. To
accomplish the escape the considered objects must cross an energetic
barrier, separating the local potential minimum from a neighboring
attracting domain. From the perspective of  statistical physics
mainly the thermally activated escape, based on the permanent
interaction of the considered system with a heat bath, has been
studied \cite{RMP}. The coupling to the heat bath causes dissipation
and local energy fluctuations and the escape process is conditioned
on the creation of a rare, optimal fluctuation which in turn
triggers an escape. To put it differently, an optimal fluctuation
transfers sufficient energy to the system so that the system is able
to statistically surmount the energetic bottleneck associated with
the transition state. Characteristic time-scales of such a process
are determined by the inverse of corresponding rates of escape out
of the domain of attraction. Within this topic, numerous extensions
of Kramers escape theory and of first passage time problems  have
been widely investigated \cite{RMP,JSPHa}. Early generalizations to
multi-dimensional systems date back to the late $1960$'s
\cite{Langer}. This method is by now well established and is
commonly put to use in biophysical contexts and for great many other
applications occurring in physics and chemistry and related areas
\cite{marchesoni}-\cite{Cattuto}.

In order that the system comprised of coupled units may pass through
a transition state an activation energy $E_{act}$ has to be
concentrated in the corresponding critical localized mode (LM). In
view of controlling the process of barrier crossing we intend to
demonstrate that the formation of the critical LM can be distinctly
accelerated via the application of a weak external ac-driving.
By use of optimally oscillating barrier configurations it is
feasible that
a far faster escape can be promoted, leading to a drastic
enhancement of the escape dynamics. Particularly at low
temperatures, where the rate of thermal barrier crossing is
exponentially suppressed, such a scenario can be very beneficial.

Prior studies mainly dealt with the appearance of LMs in damped,
driven   deterministic nonlinear lattice systems
\cite{Marin}-\cite{Maniadis}. Furthermore, the spontaneous formation
of LMs (breathers) from thermal fluctuations in lattice systems,
when thermalized  with the Nos\'e-method \cite{Nose} has been
demonstrated in \cite{Peyrard}-\cite{GTS}. Here we explain LM
formation in a stochastic system involving dissipation in presence
of enduring spatio-temporal random forcing. In addition a weak
external ac-field is applied
rendering coherently {\it oscillating barriers}.
The energy is introduced in the lattice coherently in the form of a
plane wave excitation as the response to the external ac-field and
non-coherently through thermal fluctuations. We shall demonstrate
that the stochastic source and the external ac-field conspire to
produce such an instability mechanism of the stationary flat state
(plane wave) solution yielding a spatially localized system state.
Most importantly the formed LMs prove to be robust despite the
continuously impacting thermal forces.

It should be noted that thermally activated escape of ensembles of
non-interacting (individual) particles over a metastable potential
landscape that is additionally subjected to either stochastic
 or coherent perturbations in the form of fluctuations or periodic
driving has been studied in the prior literature, e.g. see in Ref.
\cite{Doering,Lehmann,spagnolo}.  For a comprehensive overview we
refer the reader to Ref.  \cite{Reimann}. In particular, a
 resonant activation is observed, i.e.,  the mean escape
time (or the rate of escape \cite{Pechukas}) attains a minimum
(maximum)  as a function of the correlation time of the fluctuations
or the temporal driving period of the underlying potential
variations. Moreover, the kink drift motion induced by oscillating
external fields needs to be mentioned in this context
\cite{Sukstanski}. Concerning a system of coupled elements the
kink-antikink nucleation within a $\phi^4$ chain model subjected to
a deterministic periodic signal and uncorrelated noise has been
studied in \cite{AESR}. For optimal noise and coupling strength
spatiotemporal (array enhanced) stochastic resonance is observed in
the array of overdamped coupled elements. With the present study we
focus on the collective nature of the ac-driven escape process of
interacting weakly damped particles.

In detail, we study a one-dimensional lattice of damped nonlinear
and ac-driven coupled oscillators which are subjected additionally
to a heat bath at temperature $T$. Throughout the following we shall
work with dimensionless parameters, as obtained after appropriate
scaling of the corresponding physical quantities. The coordinate $q$
of each individual nonlinear oscillator with a unit mass  evolves in
a cubic, single well on-site potential of the form
\begin{equation}
U(q)=\frac{\omega_0^2}{2}q^2-\frac{a}{3}q^3.
\end{equation}
This potential possesses a metastable equilibrium at $q_{min}=0$,
corresponding to the rest energy $E_{min}=0$ and exhibits a maximum
that is located at $q_{max}=\omega_0^2/a$ with energy $E_{max}\equiv
\Delta E=\omega_0 ^6/(6 a ^2)$. Thus, in order for particles to
escape from the potential well of depth $\Delta E$ over the energy
barrier and subsequently into the range $q>q_{max}$, a sufficient
amount of energy need to be supplied. The lattice dynamics is
governed by the following system of coupled Langevin equations
\begin{eqnarray}
\lefteqn{\ddot{q}_n+\gamma \dot{q}_n+\omega_0^2 q_n-a q_n^2+\xi_n(t)}\nonumber\\&-&\kappa
\left[q_{n+1}+q_{n-1}-2q_n\right]
-f\,\sin(\omega t+\theta_0)=0\,.\label{eq:qdot}
\end{eqnarray}
The coordinates $q_{n}(t)$ quantify the displacement of the
oscillator in the local on-site potential $U$ at lattice site $n\in
[1,N]$. The oscillators,  referred to as "units", are coupled
bi-linearly to their neighbors with interaction strength $\kappa$.
The friction strength is measured by the parameter $\gamma$ and
$\xi_n(t)$ denotes a Gaussian distributed thermal, white noise of
vanishing mean $\langle\xi_n(t)\rangle=0$, obeying the well-known
fluctuation-dissipation relation
\begin{equation}
\langle\xi_n(t) \xi_{n^{\prime}}(t^{\prime})\rangle=2\gamma k_B
T\delta_{n,n^{\prime}}\delta(t-t^{\prime}) \;,
\end{equation}
with $k_B$ denoting the Boltzmann constant. A homogeneous external
periodic modulation field of amplitude $f$, frequency $\omega$ and
phase $\theta_0$ globally acts upon the system. In this work we use
periodic boundary conditions according to $q_{N+1}=q_1$ and fix the
parameters of the  potential as follows: $\omega_0^2=2$ and $a=1$,
yielding $\Delta E = 4/3$. A deterministic escape scenario in the
conservative, undriven limit of system (\ref{eq:qdot}) has been
explored by us in \cite{EPL,PRE07}.

\begin{figure}
\begin{center}
\includegraphics[scale=0.38]{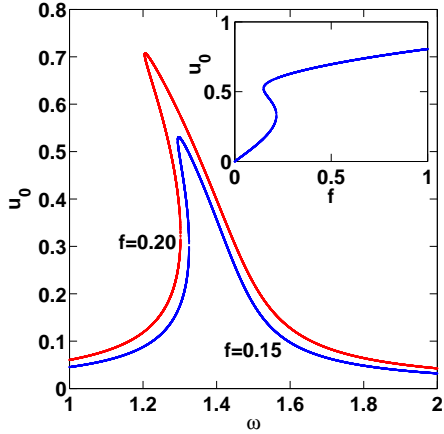}
\caption{\label{fig:omega-u0}(color online) Response of the
amplitude $u_0$ with respect to the frequency $\omega$ of the
driving force for two fixed values of the driving amplitude as
indicated in the plot. The damping constant is $\gamma=0.1$. The
inset shows the response of the amplitude $u_0$ with respect to
the driving amplitude $f$ for a fixed frequency $\omega=1.295$.  The remaining parameter values are $\omega_0^2=2$ and $a=1$.}
\end{center}
\end{figure}

\begin{figure}
\begin{center}
\includegraphics[scale=0.38] {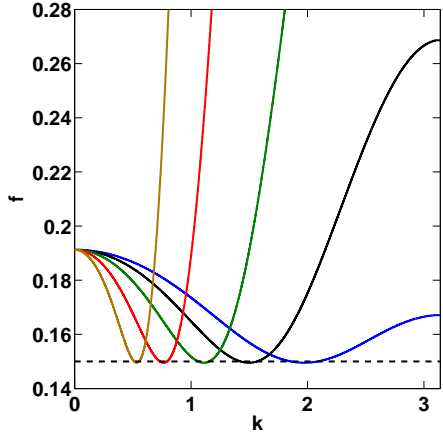}
\caption{\label{fig:tongue}(color online) Instability bands for
different coupling strengths $\kappa=2,1,0.5,0.3,0.2$ (decreasing
from left to right) are displayed. The relation between the
amplitude of the external ac-field $f$ and the amplitude of the
homogeneous solution $u_0$ is the one displayed with  the inset in
Fig.~\ref{fig:omega-u0}. The horizontal dashed line at $f=0.15$
intersects each instability band very close to its bottom, the
position of which determines the respective critical wave number
$k_c$ (see also text).}
\end{center}
\end{figure}

\begin{figure}
\begin{center}
\includegraphics[scale=0.44]{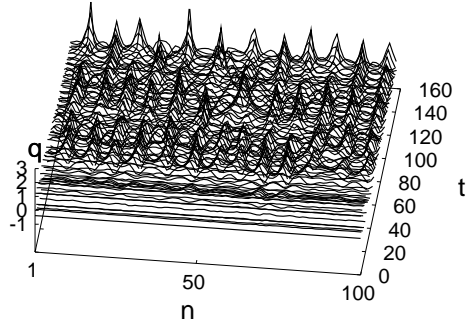}
\includegraphics[scale=0.44]{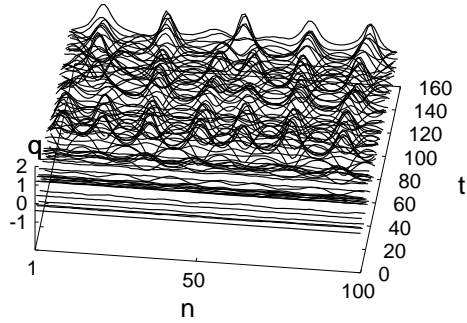}
\includegraphics[scale=0.44]{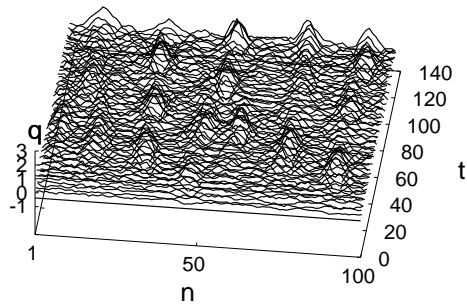}
\caption{\label{fig:pattern} Spatio-temporal pattern
of the solutions $q_n(t)$ for a lattice consisting of $N=100$ sites
and chosen coupling strength at $\kappa=0.5$ (top panel), $\kappa=2$
(central panel) for the same realization of Gaussian white noise
with thermal energy $k_BT=0.001\times \Delta E$. Bottom panel: Same
as in the central panel but now for a $50$-times larger thermal
energy: $k_BT=0.05\times \Delta E$. The remaining parameter values
are set at $f=0.15$, $\omega=1.295$, $\theta_0=0$ and friction
$\gamma=0.1$.}
\end{center}
\end{figure}

To analyze the nonlinear character of the solutions of
Eq.\,(\ref{eq:qdot}) we first discard the noise ($\xi_n=0$) and
derive a nonlinear damped and driven discrete Schr\"{o}dinger
equation for the slowly varying envelope solution, $u_n(t)$,
following the reasoning in \cite{Kivshar93}, i.e.,
\begin{eqnarray}
\lefteqn{2i\omega_0\,\dot{u}_n+i\gamma \omega_0 u_n
+\kappa\,\left[u_{n+1}+u_{n-1}-2u_n\right]}\nonumber\\
&+&\alpha\,|u_n|^2 u_n+\frac{1}{2}f\,\exp[-i\Delta \omega t+\theta_0]=0\,,\label{eq:DNLS}
\end{eqnarray}
with the nonlinearity parameter reading
$\alpha={10a^2}/{3\omega_0^2}$ and $\Delta \omega =\omega-\omega_0$.
For the amplitude $u_0$ of a spatially homogeneous solution of
Eq.\,(\ref{eq:DNLS}) of the form
\begin{equation}
u_n(t)=u_0\,\exp[-i(\Delta \omega
t+\theta_0)]+c.c.
\end{equation}
one obtains
\begin{small}
\begin{equation}
\left[\,\left(2\omega_0 \Delta \omega +
\alpha u_0^2\right)^2+\gamma^2 \omega_0^2\,\right] u_0^2=\frac{1}{4}f^2\,.\label{eq:response}
\end{equation}
\end{small}
In Fig.~\ref{fig:omega-u0}
we depict the amplitude $u_0$ of the response versus the driving
frequency curve for two different values of the driving amplitude
$f$. At a bifurcation point a "jump" resonance related with a
saddle-node bifurcation occurs and in certain range of the driving
frequency multistability exists. In comparison for the larger
driving amplitude, $f=0.2$, the bifurcation point for the "jump"
resonance occurs at a lower frequency value than for the driving
with $f=0.15$. Moreover, in the former case the system responds
overall with higher amplitudes $u_0$ than in the latter. Notice that
the system responds with large amplitude only within a frequency
window and large amplitudes are obtained for a driving frequency
lying below the band of linear frequencies, viz.
$\omega<\omega_0=\sqrt {2} =1.414...$. Similarly, for the response
of the amplitude with regard to the driving strength $f$ multistable
solutions are possible as depicted with the inset in
Fig.~\ref{fig:omega-u0}. In order to investigate the stability of
the homogeneous solution of Eq.~(\ref{eq:qdot}) we use
\begin{eqnarray}
q_n(t; k=0)&=&x(t)=u_0e^{-i(\omega
t+\theta_0)}\nonumber\\
&+&\frac{a}{\omega_0^2}\left[2-\frac{1}{3} e^{-2i(\omega
t+\theta_0)}\right]u_0^2+c.c. \; .\label{eq:homogeneous}
\end{eqnarray}
and write with respect to the spatial perturbations $A_n$:
$q_n(t)=x(t)+A_{n}(t)$. Since we impose periodic boundary conditions
the Fourier-series expansion $A_n(t)=\sum_k\,\exp(ikn)s_k(t)$ can be
used to yield an equation for the mode amplitudes $s_k$, i.e.,
\begin{equation}
 \ddot{s}_k+\gamma \dot{s}_k +\omega_k^2 s_k-4 a u_0\cos(\omega t+\theta_0)s_k=0\, ,
\end{equation}
where we discarded a higher harmonics  and introduced
$\omega_k^2=\omega_0^2+4\kappa \sin^2\left({k}/{2}\right) -8\left({a
u_0}/{\omega_0}\right)^2$. Setting $\tau=\omega t/2$ and
$s_k(t)=v_k(t)\exp(-\gamma t/2)$ one derives a Mathieu equation
\begin{equation}
\ddot{v}_k+[A-2Q \cos(2t+2\theta_0)]v_k=0\,,
\end{equation}
with the parameters $A=(2\omega_k/\omega)^2 -(\gamma/\omega)^2$ and
$Q=8 a u_0/\omega^2$. If it holds that $\sqrt{A}\simeq l$, with $l$
denoting a positive integer number, the Mathieu equation allows for
parametric resonance \cite{Arnold},\cite{Landau}.
The extension of
the resonance regions is determined by the ratio $Q/A$;  for the
primary resonance, $A\simeq 1$, it is given by
\begin{equation}
(A-1)^2<Q^2.
\end{equation}
For the parameter set corresponding to the line shown in the inset
in Fig.~\ref{fig:omega-u0} (determining the relationship between the
amplitude $f$ of the external ac-field and the amplitude of the
homogeneous solution $u_0$) the instability bands for different
values of the coupling strength  $\kappa$ are depicted in
Fig.~\ref{fig:tongue}.
For the onset of parametric resonance the driving amplitude $f$
 has to exceed the value of the  bifurcation
point, i.e. $f_c\gtrsim 0.1408$ related with the "jump" resonance,
regardless of the value of $\kappa$. The position of the bottom of
the instability band, determining the critical unstable wave
number $k_c$, shifts towards  lower $k-$values with increasing
coupling strength $\kappa$. For a chosen field strength $f=0.15$,
that lies just above $f_c$, one expects that the LMs of distinct
wave length, determined by
$\lambda_c=2\pi /k_c$ become excited (cf. Fig.~\ref{fig:tongue}).
We infer from Fig.~\ref{fig:tongue} that the wave length of a LM
increases with increasing coupling $\kappa$. This is verified in
Fig.~\ref{fig:pattern}
showing the spatio-temporal evolution of the amplitudes $q_n(t)$
for couplings $\kappa=0.5$ and $\kappa=2$. The Langevin equations
were numerically integrated using a two-step Heun stochastic
solver. In all our simulations the initial chain configuration is
represented by  $q_n(0)=x(0)$ and $p_n(0)=0$ with the homogeneous
solution (plane wave) $x(0)$ given in (\ref{eq:homogeneous}).
We note the formation of a LM of certain wave length arising from
the homogeneous state after a short time span (after $t\sim 60$)
and we note that the period duration for oscillations near the
bottom of the potential is around $2\pi/\omega_0\simeq 4.4$.

Regarding the energy relation within the stationary flat state
(where each unit contains the same amount of initial energy) we
monitored the temporal evolution of the energy of one unit
\begin{equation}
E_{n}=\frac{1}{2}
p_n^2+U(q_n)\,,
\end{equation}
and the corresponding field energy
\begin{equation}
E_{field}=-f \sin (\omega t+ \theta_0 )q_n\,,
\end{equation}
without coupling the chain to the heat bath for a force amplitude
$f=0.15$. In this stationary case the field energy performs
small-amplitude oscillations around a mean value of
${E}_{field}=0.04 \equiv 0.03\times \Delta E= 0.03 \times 4/3$,
while the mean of the energy of one unit is $E_n=0.78=0.585\times
\Delta E$ (not shown). Thus the gain of energy, determined by the
ratio $E_{n}/E_{field}$, amounts to a remarkable high value of $19.5$. 
To retain this relation upon lowering (increasing) the
damping $\gamma$ a lower (higher) driving strength $f$ is
necessary while the "jump" resonance frequency attains a lower (higher) value according to Eq.~(\ref{eq:response}).

The stochastic term provides perturbations of all wave numbers and
a pattern emerges from the homogeneous flat state. That is, due to
the effect of parametric resonance perturbations provided by the
thermal noise grow and induce a LM consisting of several humps.
The fastest growing perturbations are those associated with the
critical wave number $k_c$ (see also \cite{Kolomeisky}). Each of
these humps resembles the hairpin shape of the transition state as
the critical escape configuration possessing an  energy  $E_{act}$
through which the coupled units have to pass in order to cross the
barrier \cite{PRE07}. The robustness of the LMs is remarkable: a
LM is sustained despite continuously impacting thermal noise of
strengths  up to values $k_BT\lesssim 0.2\times \Delta E$.
Moreover, the formed pattern maintain their distinct wave length  $\lambda_c=2\pi/k_c$
(see Fig.~\ref{fig:pattern}).

\begin{figure}
\begin{center}
\includegraphics[scale=0.38]{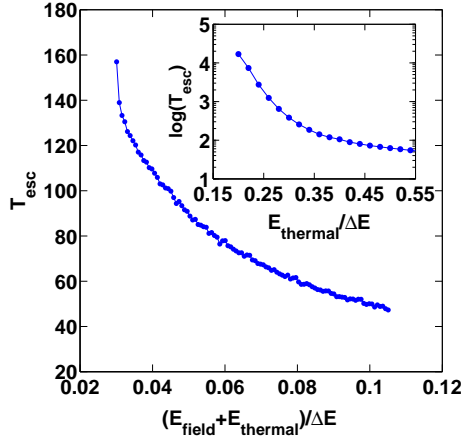}
\caption{\label{fig:tesc} The mean escape time of the chain versus
the mean injected energy $E={E}_{field}+ k_B T$ measured in units of
$\Delta E$ with {\it fixed} field energy ${E}_{field}=0.03\times
\Delta E$ provided by an external modulation field with
$\omega=1.295$, $\theta_0=0$ and $f=0.15$. Here we vary  the thermal
energy $E_{thermal}=k_BT$. The inset depicts the unforced case with
$f=0$. The remaining parameter values are $N=100$, $\kappa=0.28$ and
$\gamma=0.1$.}
\end{center}
\end{figure}

We note that upon increasing the noise strength  the growth rate of
the humps becomes enhanced, being  reflected in the statistics of
the barrier crossing of the chain in the presence of weak
ac-driving. The amplitude and frequency of the latter are chosen
such that the dynamics exhibits parametric resonance. The dependence
of the mean escape time of the chain on the injected average energy
$E \equiv E_{field}+E_{thermal}$, with $E_{thermal} \equiv k_B T $
(measured in units of the barrier energy $\Delta E$) is displayed in
Fig.~\ref{fig:tesc}. The thermal energy $E_{thermal}$, supplied non-coherently by the
heat bath, is {\it varied} within the range
$[(10^{-4}-0.11)\times\Delta E]$.

The average of the escape times was performed over $500$
realizations of the thermal noise. In this context the random escape
time of a unit is defined as the time instant when the unit passes
through the value $q=20$ far beyond the potential barrier. Thus, no
likely recrossing back into the potential valley can occur
\cite{EPL,PRE07}. The escape time of the chain is then determined by
the average of the escape times of its units. We notice that the underlying irregular dynamics serves for
self-averaging  and thus the choice of the phase of the coherent,
external forcing, $\theta_0$, does not affect the mean escape time. In the forced as well as unforced
case there occurs a rather rapid decay of $T_{esc}$ with growing
$E_{thermal}=k_BT$ at low temperatures. This effect weakens
gradually upon further increasing $k_BT$. Most strikingly, for the
forced system the escape times become drastically shortened in
comparison with the unforced case with $f=0$. Moreover, for the
forced system escape takes place also at very low temperatures for
which in the undriven case  not even the escape of a single unit has
been observed during the simulation time (taken here as $t=10^5$)
implying a {\it giant enhancement of the rate of escape} as compared
to the purely thermal noise driven rate.

\begin{figure}
\begin{center}
\includegraphics[scale=0.38]{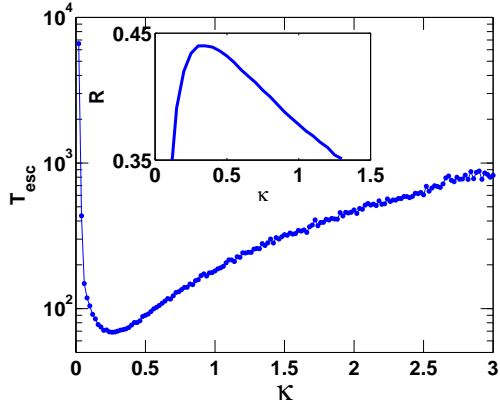}
\caption{\label{fig:kappa-tesc} The mean escape time versus the coupling strength $\kappa$ exhibits a resonance structure. The inset displays the ratio $R$, defined in Eq.~(\ref{eq:r}), as a function of $\kappa$. The remaining parameter values are given by
$N=100$, $k_BT=0.05\times \Delta E$, $f=0.15$, $\omega=1.295$,
$\theta_0=0$ and $\gamma=0.1$.}
\end{center}
\end{figure}

Upon exploring the optimal escape route  we investigated  the
influence of the coupling strength $\kappa$ on the average escape
time. Our numerical findings are summarized in
Fig.~\ref{fig:kappa-tesc}.
The mean escape time exhibits a {\it resonance structure}, viz. there exists an optimal  coupling strength ($\kappa_{res}\simeq
0.28$) for which the escape assumes a minimum. Upon lowering
$\kappa <\kappa_{res}$ we notice a drastic rise of the escape time
while for $\kappa > \kappa_{res}$ the graph exhibits only a
moderately growing slope with growing coupling strength $\kappa$.
We emphasize  the {\it collective} nature of this resonance effect
which here occurs for {\it finite} interaction strength $\kappa
\neq 0$. In the limit  $\kappa \rightarrow 0$ the mean escape time
of noninteracting,  individual particles assumes for this
parameter set an extreme large value, implying a vanishingly small
escape rate.

To explain the occurrence of the resonance structure in
Fig.~\ref{fig:kappa-tesc} we recall that the wave length,
$\lambda_{c}=2\pi/k_c$, of the arising LMs on the lattice is
determined by the critical wave number $k_c=k_c(\kappa,f)$ (cf.
Fig.~\ref{fig:tongue}). The number of humps contained in a LM,
$N_{h}$, can be attributed to $k_c$ as: $\lambda_{c}N_{h}=2\pi/k_c
N_h=N$. The number of humps (besides their height and width)
regulates how the mean energy injected via the coherent  external
field and the incoherent thermal noise is shared among them.
Supposing that the whole lattice can be divided into an array of
segments, where each of them supports a single localized hump, the
energy of one segment is given by
$E_s=E/N_{h}=2\pi E/(k_c N)$. Appropriate conditions for successful escape are provided when the
energy contained in each segment, $E_s$, is close to the
activation energy, $E_{act}$ of the critical escape configuration \cite{PRE07}. The efficiency of energy localization is then determined by the ratio
\begin{equation}
R={E_s}/{E_{act}}.\label{eq:r}
\end{equation}

\begin{figure}
\begin{center}
\includegraphics[scale=0.38]{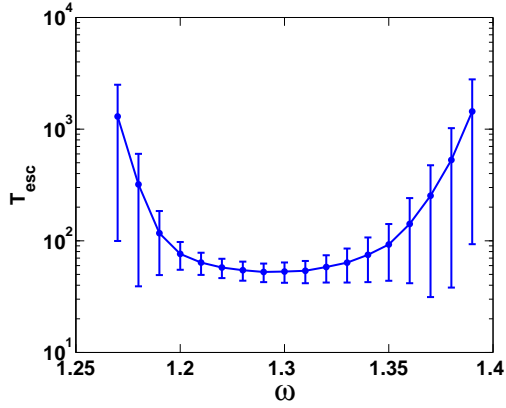}
\caption{\label{fig:resonance} The mean escape time as a function
of the driving frequency. The thermal energy is $k_BT=0.05\times
\Delta E$ and the driving amplitude is $f=0.15$. The remaining
parameter values are $N=100$, $\omega=1.295$, $\kappa=0.3$,
$f=0.15$ and $\gamma=0.1$. For comparison we note that Langer's
theory \cite{Langer} yields for the mean escape time in the
unforced case, $f=0$, the value $T_{esc}\simeq 553\times 10^5$
underpinning the drastic speed-up effect in our forced case.}
\end{center}
\end{figure}

The activation energy as a function of the coupling strength
satisfies (we recall that we use a dimensionless formulation) the relation $E_{act}=(1+3.54\times\kappa)\Delta E$ \cite{PRE07}.
Keeping the injected energy $E$  fixed and given value of $k_c$ we
obtain $R$. In the inset of Fig.~\ref{fig:kappa-tesc} the ratio
$R$ is plotted as a function of the coupling strength $\kappa$.
The plot indeed exhibits a maximum at $\kappa=0.28$, which
confirms the finding of the resonance found for the mean escape
time versus coupling strength as depicted in
Fig.~\ref{fig:kappa-tesc}. 
Concerning the critical localized mode through which a lattice
state has to pass through in order to escape over the potential
barrier we remark that for comparatively low coupling
strengths ($\kappa \lesssim 0.6$) the effect of discreteness in
the lattice system is still such pronounced that this critical
localized mode is indeed represented by a thin hairpin-shaped
configuration involving effectively one lattice unit of large
amplitude to either side of which the amplitude pattern decays
extremely rapidly (for more details see \cite{PRE07}).

Next we study the role of the angular driving   frequency of the
external modulation field, see in Fig.~\ref{fig:resonance}. The
escape time as a function of the angular frequency likewise exhibits
a resonance structure and there exists  an optimal  frequency for
which the average escape time assumes a minimum.
This is reminiscent of the phenomenon of resonant activation found
for the thermally activated escape of noninteracting particles
surmounting oscillating barriers \cite{Doering, Lehmann, spagnolo,
Reimann, Pechukas}.
In our case of a nonlinear chain composed of coupled units, however,
this 'resonant activation' within a frequency window nicely
correlates with the  systems' gain of energy that is coherently
supplied by the applied ac-field in this very same frequency
interval (note the corresponding frequency window in the nonlinear
frequency response graph associated with large amplitudes in
Fig.~\ref{fig:omega-u0}). Therefore, tuning the frequency $\omega$
at a fixed interaction strength $\kappa$ allows to optimize further
the mean escape time. The minimal escape scenario thus requires an
optimal tuning  both in coupling strength and ac-driving frequency
$\omega$.

In summary, we have presented a drastic speed-up mechanism of the
thermal  noise driven barrier crossing of a coupled damped nonlinear
oscillator chain under the impact of a weak external ac-field. With
appropriate parameter values of the latter and in the presence of
thermal noise an instability mechanism is initiated due to which LMs
arise from stationary flat state solutions in the lattice dynamics.
Humps of the LMs are rapidly driven through the transition state
thus accelerating the escape over the situation with purely
thermally assisted escape. Interestingly, the LMs are sustained up
to fairly high noise levels corresponding to $k_BT\simeq 0.2\times
\Delta E$.

The findings of our study can be applied for the control of the rate
of barrier crossing of oscillator chains. With such chains providing
the archetype model for nonlinear collective transport  of matter,
charge and energy in abundant low-dimensional systems in physics,
biology and chemistry this speed-up scenario of thermally driven
collective escape over potential barriers might well be put to
constructive use in a variety of potential applications.

\vspace{0.2cm}

\centerline{\large{\bf Acknowledgments}}
\noindent This research has been supported by SFB-555 (L. Sch.-G) and,
as well, by  the joint Volkswagen Foundation projects I/80424 (P. H.) and I/80425 (L. Sch.-G).


\begin{thebibliography}{99}
\bibitem{RMP} H\"anggi P., Talkner P. and Borkovec M., Rev.~Mod.~Phys. {\bf 62} (1990) 251.
\bibitem{JSPHa}
H\"anggi P., J.~Stat.~Phys. {\bf 42} (1986) 105; J.~Stat.~Phys.
\textbf{44} (1986) 1003 .
\bibitem{Langer} Langer J.~S., Ann. Phys. (N.Y.) {\bf 54} (1969) 258 .
\bibitem{marchesoni}
H\"anggi P., Marchesoni F., and Sodano P., Phys. Rev. Lett. {\bf
60} (1988) 2563; H\"anggi P. and Marchesoni F., Phys. Rev. Lett.
{\bf 77} (1996) 787.
\bibitem{AESR} Marchesoni F., Gammaitoni L., and Bulsara A.R., Phys. Rev. Lett. {\bf 76} (1996) 2609.
\bibitem{Sung} Sung W. and Park P.~J., Phys. Rev. Lett.{\bf 77} (1996) 783.
\bibitem{Park} Park P.~J. and  Sung W., Phys. Rev. E {\bf 57} (1998) 730 ; J. Chem. Phys.
{\bf 108} (1998) 3013 ; {\it ibid} {\bf 111} (1999) 5259.
\bibitem{Sebastian} Sebastian K.~L. and Paul A.~K.~R., Phys. Rev. E {\bf 62} (2000) 927.
\bibitem{Lee}
Lee S. and Sung W., Phys.~Rev.~E {\bf 63} (2001) 021115; Lee K.
and Sung W., Phys. Rev. E {\bf 64} (2001) 041801.
\bibitem{Kraikivsky} Kraikivsky P., Lipowsky R., and Kiefeld J.,
  Europhys. Lett. {\bf 66} (2004) 763 .
\bibitem{Linke}  Dowtown M.~T., Zuckermann M.~J.,  Craig E.~M., Plischke M., and Linke H., Phys. Rev. E {\bf 73} (2006) 011909.
\bibitem{BM} H\"anggi P., Marchesoni F., and Nori F., Ann.~Physik (Leizig)
\textbf{14} (2005) 51.
\bibitem{astumian} Astumian R.~D. and H\"anggi P., Physics
Today {\bf 55} (11) (2002) 33; Reimann P. and H\"anggi P., Appl.~Phys. A {\bf 75} (2002) 169.
\bibitem{Cattuto} Cattuto C. and Marchesoni F., Europhys. Lett. {\bf 62} (2003) 363.
\bibitem{Marin} Marin J.L. and Aubry S., Nonlinearity {\bf 9} (1994) 1501.
\bibitem{Hennig98} Hennig D., Phys. Rev. E {\bf 59} (1998) 1637.
\bibitem{Vanossi} Vanossi A., Rasmussen K.\O., Bishop A.R., Malomed B.A., and Bortolani V., Phys. Rev. E {\bf 62} (2000) 7353.
\bibitem{Marin1} Marin J.L., Falo F., Martinez P.J., and Floria L.M., Phys. Rev. E {\bf 63} (2001) 066603.
\bibitem{Maniadis}
Maniadis P. and Flach S., Europhys. Lett. {\bf 74} (2006)  452.
\bibitem{Nose}  Nos\'e S., J. Chem. Phys. {\bf 81} (1984) 511.
\bibitem{Peyrard} Peyrard M., Physica D {\bf 119} (1998) 84; Dauxois T., Peyrard M., and Bishop A.R., Phys. Rev. E {\bf 47} (1993) 684.
\bibitem{GTS} Tsironis G.P. and Aubry S., Phys. Rev. Lett. {\bf 77} (1996) 5225.
\bibitem{Doering} Doering C. and Gadoua J.C., Phys. Rev. Lett. {\bf 69} (1992) 2318.
\bibitem{Lehmann} Lehmann J., Reimann P., and H\"anggi P., Phys. Rev. Lett {\bf 84} (1999) 1639; Phys. Rev. E {\bf 62} (2000) 6282.
\bibitem{spagnolo}
Mantegna R.~N. and Spagnolo B., Phys. Rev. Lett. {\bf 84} (2000) 3025.
\bibitem{Reimann} Reimann P. and H\"anggi P., {\it Surmounting fluctuating
barriers: Basic concepts and exact results};  in:  Stochastic
Dynamics, eds. L. Schimansky-Geier and Th. P\"oschel, Lecture Notes
in Physics, {\bf 484} (1997) 127.
\bibitem{Pechukas} Pechukas P. and H\"anggi P., Phys. Rev. Lett. {\bf 73} (2004) 2772.
\bibitem{Sukstanski} Sustanski A.L. and Primak K.I., Phys. Rev. Lett. {\bf 75} (1995) 3029; Kivshar Yu.~S. and Sanchez A., Phys. Rev. Lett. {\bf 77} (1996) 582.
\bibitem{EPL} Hennig D., Schimansky-Geier L., and  H\"{a}nggi P., Europhys. Lett. {\bf 78} (2007) 20002.
\bibitem{PRE07} Hennig D., Fugmann S., Schimansky-Geier L., and H\"{a}nggi P. Phys. Rev. E {\bf 76} (2007) 041110.
\bibitem{Kivshar93}  Kivshar Yu. S., Phys. Rev. E {\bf 48} (1993) 4132; Daumont I., Dauxois T., and Peyrard M., Nonlinearity {\bf 10} (1997) 617.
\bibitem{Arnold} V.I. Arnold, {\em Mathematical Methods of Classical Mechanis} (Springer, New York, 1997).
\bibitem{Landau}  L.D. Landau and E.M. Lifshitz {\em Course of Theoretical Physics: Mechanics} (Akademie-Verlag, Berlin, 1987).
\bibitem{Kolomeisky} Kolomeisky E.B., Curcic T, and Straley J.P., Phys. Rev. Lett. {\bf 75} (1995) 1775.
\end{thebibliography}
\end{document}